# Modelling of limitations of bulk heterojunction architecture in organic solar cells

*Jacek Wojtkiewicz[1] and Marek Pilch[2]*
*Faculty of Physics, Warsaw University*
*ul. Pasteura 5, 02-093 Warszawa, Poland*
[1]*e-mail: wjacek@fuw.edu.pl,* [2]*e-mail: Marek.Pilch@fuw.edu.pl*

**Abstract:**

Polymer solar cells are considered as very promising candidates for development of photovoltaics of the future. They are cheap and easy to fabricate, however, up to now, they possess fundamental drawback – low effectiveness. In the most popular BHJ (bulk heterojunction) architecture the actual record of efficiency is about 13%. One ask the question how fundamental this limitation is. In our paper we propose the simple model which examines the limitations of efficiency by analysis of geometrical aspects of the BHJ architecture. In this paper we considered two-dimensional model. We calculated the effective length of the donor-acceptor border in the random mixture of donor and acceptor nanocrystals and further compared it with an ideal 'comb' architecture. It turns out that in the BHJ architecture, this effective length is about 2 times smaller than in the 'comb' architecture.

1. Introduction

Organic photovoltaics is considered as one of the most perspective investigational trends in entire topic of new types of solar cells design. Main advantages of the organic photovoltaic cells are: low cost, flexibility and small weight. Regrettably, the price we have to pay so far is low effectiveness: for few years the record has been fixed on level of 12% [1]. One can ask what are perspectives to improve this effectiveness to at least commercial silicon cells performance (18-24 % [1]). Related questions have been posed many times [3], [18].

In order to recognize various aspects of the problem, let us first remind the basic mechanism of the action of solar cell. The conversion of light into electric current in organic cell is a complex, multistage process. One can recognize the following main stages of it [2], [3].

Basic elements of active layer of a cell are: the electron donor and the acceptor. In most cases, the donor is an organic polymer or oligomer. On the other hand, fullerenes or their chemical derivatives are used in most cases as acceptors. In the first stage, the donor absorbs photons of solar light. After absorption, an exciton is formed (i.e. a bound state of excited electron and a hole). It diffuses to the border between donor and acceptor. On the border, the dissociation of an exciton into an electron and a hole takes place. The hole remains in the donor, and the electron moves to the acceptor. Next, the carriers of electric charge wander to the electrodes, where they accumulate. As a result we observe the voltage between the electrodes.

An opportunity, which must be taken into account during solar cells designing is a short diffusion length of an exciton. In most cases, it is of the order of a few nanometers, rarely exceeding this value to about 20-30 nm [4]. Historically, the first layer solar cells have been built in a simple layer architecture (Fig. 1), where thickness of the layers were 50-100 nm [4-8]. It turns out that they give limited effectiveness (up to 5%). The most important factor which limits their efficiency is that majority of the excitons decays before they achieve the border with an acceptor.

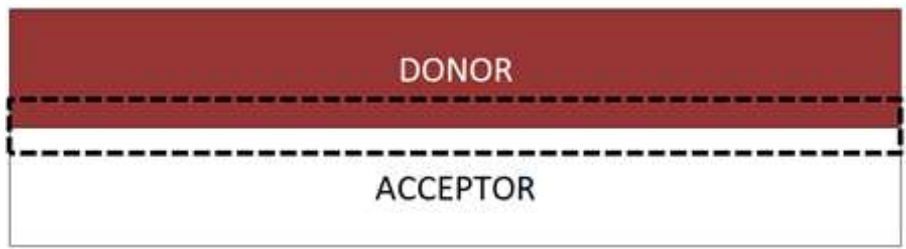

Fig. 1: Schematic view of the layer architecture.

Partial solution of this problem is given by the most popular now architecture called BHJ (Bulk HeteroJunction). [9-12] (Fig. 2a). Here, the active layer represents compound seeds of donor and acceptor with characteristic scale tens of nanometers. This makes the area of D-A contact to be large and an exciton can get the border with an acceptor with high probability. It's great opportunity of the BHJ architecture. The another opportunity is it's simplicity: to prepare an active BHJ blend, it suffices to mix donor's and acceptor's solutions and after evaporation the solvent, the blend is ready.

a)

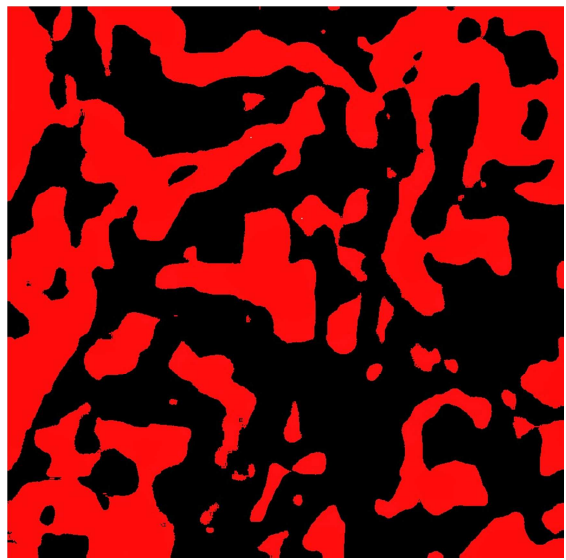

b)

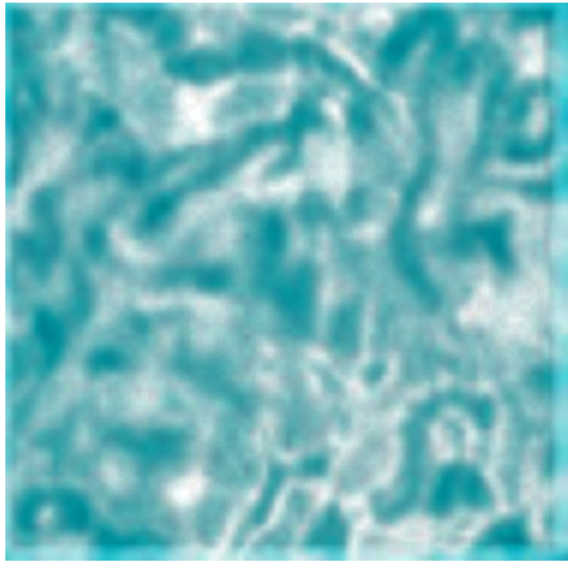

c)

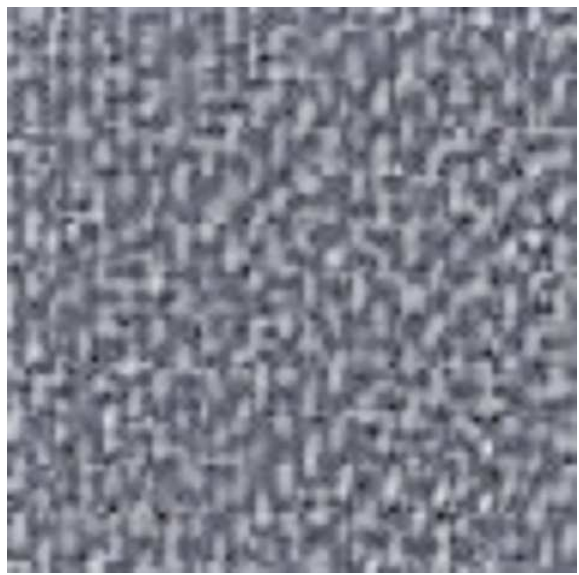

Fig. 2. a) Very schematic view of the BHJ architecture; b),c) – views of the BHJ architecture supported by TEM images

However, BHJ architecture has also certain drawbacks. One of them is a creation of the 'islands' of donor and acceptor i.e. attendance of the seeds, which have no connection with any electrodes. In such a situation, even if a charge is generated on an 'island', it can't go to the proper electrod. This mean losses in the cell's effectiveness. Analogical losses are caused by attendance of the 'bad peninsulas', i.e. the donor's seeds against the cathode and the acceptor's seeds against the anode. It's obvious that there are factors affecting negatively the solar cell's effectiveness in the BHJ architecture. These negative factors were recognized very long ago, but surprisingly we couldn't find the estimate of the scale of these effects in literature.

The architecture which is well-fitted to the exciton's features is so-called 'comb architecture' (Fig. 3) [13]. Size of the donor's insets should fit the exciton's diffusion length, i.e. their characteristic width should be of the order 10-20 nm. Such devices in principle seem to be available on laboratory scale, but the fabrication of well-controlled size combs in large-scale technology is another matter. Nevertheless, comparison of effectivenesses of the optimal comb architecture versus those of the BHJ architecture seems very interesting. We couldn't find such comparisons in literature. This opportunity led us to pose the following problem: *Propose a model – even crude and simplified - which could estimate the losses due to presence of the 'islands' and the 'bad pennisulas', and farther to compare potential efficiencies of the cells in BHJ and comb architecture.*

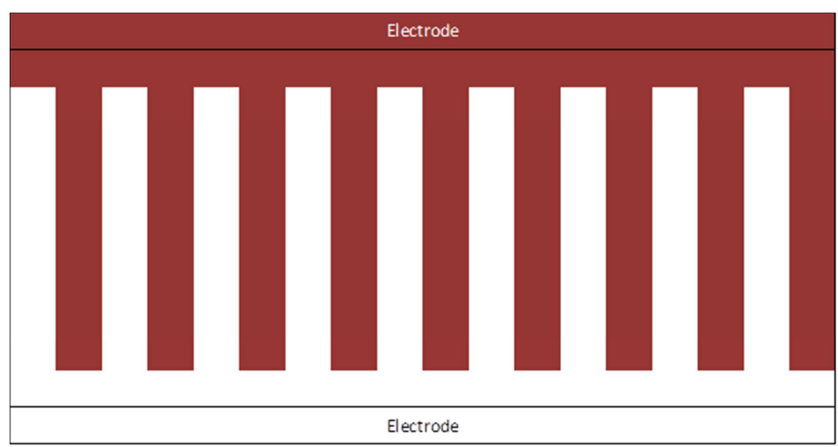

Fig. 3. Schematic view of the 'comb' architecture

Our first step is consideration of two-dimensional situation. Three-dimensional model would be much more realistic, but it's more complicated too. We plan to analyse this model in the future. We believe that the analysis of two-dimensional model is worth doing and that this is an important introductory step before the analysis of three-dimensional model.

The general setup of the model is as follows. We treat the donor's and acceptor's nanoseeds as the squares in a square lattice. We consider also the second version of the model where seeds are hexagons occupying cells of honeycomb lattice. In both versions, we assume that the donor's and acceptor's seeds are randomly distributed. Afterwards, we have computed length of the borderline between the donor's and acceptor's seeds allowing of the 'parasitic' effects attendance, i.e. fact that the 'islands' and the 'bad pennisulas' borderlines contribute nothing to production of electricity. This way, so-called 'effective' length of the D-A borderline was calculated. In the next step, we compared it with the length of the borderline in the comb architecture. As a result, *it turned out that in the average the 'active' length of the borderline in comb architecture was 2-2.5 times bigger than their counterpart in BHJ.*

The organization of the paper is as follows. In the Sec. 2 we lay out assumptions of the model and present the simulation's algorithm. In the Sec. 3, we present in some details results obtained. The Sec. 4 is devoted to summary; perspectives of further investigations are also sketched.

**2. The model and the computational algorithm**

**2.1. Kinds of blends which we simulate**

Fig. 2a is a *schematic* illustration of a BHJ blend. Very similar figures can be found in numerous papers, see for instance [15], Fig. 1B, where the three-dimensional version is presented. However, one can pose the question: *How such schematic figures are related to reality?* To settle this question, it is necessary to invoke experimental results. Experimental probing of the BHJ blend structure were performed in numerous papers; an exhaustive review is [16]. It turns out that various methods, for instance TEM (Transmission Electron Microscopy) suggest rather different pictures of the structure of BHJ blend. First of all, they are quite diverse. One of situations encountered is illustrated on Fig. 2b : Regions occupied by the donor and acceptor form irregular shapes, and borders between them are fuzzy. For an example of the TEM image similar to Fig. 2b , see Fig. 3D in [15]. One encounters also the 'grain' structures, similar to those presented on Fig. 2c : Regions occupied by the donor and acceptor possess similar size and boundaries between them are relatively sharp. Examples of such structures are presented for instance in [15], Figs. 3A,B,C.

Our modeling refers to the similar structures, where regions occupied by the donor and acceptor have similar size and shape, and the boundary between them not too fuzzy. We make two simplifying assumptions: *The donor and acceptor grains possess the same shape and size* (they are squares or hexagons).

Moreover, we assume that *borders between donor and acceptor domains are sharp*.

**2.2. Basic technical assumptions**:

- The model is two-dimensional one.

- We considered two variants, defined on lattices: square one (version 1) and honeycomb one (version 2).

- We assume that in version 1 the model of BHJ layer consists of randomly colored white (donor) and red (acceptor) squares (Fig. 4). Their shape and size correspond to the grain size in some real blends, which is of the order 10 nm (see for instance [4],[15], [16]). Of course, one should realize that shapes of grains in real blends can possess very different shapes and sizes. However, there is a quite large group of blends where seeds form regular (round or square) shapes of approximately the same size [15]. We took the value of grain size as 10 nm. We assume that every created exciton achieves the Donor-Acceptor boundary. In version 2, the model of BHJ layer consists of randomly colored hexagons in the honeycomb lattice. The diameter of a hexagon corresponds to grain size 10 nm.

- In version 1 were considered subsets of square lattice with sizes ranging from 10 x 10 till 20 x 200 lattice units. Length of shorter side correspondes to thickness of the BHJ layer, i.e. distance between electrodes. We adopted it as 10, 15 and 20 squares; this way cells with thickness of the BHJ layer of 100-200 nm were simulated. The another side of the BHJ layer corresponds to characteristic size of an electrode; in a real cell it is widely bigger than thickness of the BHJ layer. We assumed that the second side is 1-10 times longer than the first one (the last numer is widely close to reality). In the version 2 we adopted analogical size of subsets of the honeycomb. We present sample configurations on square and honeycomb lattices on Figs. 4 and 6, respectively.

- In cells we can have various ratio of donor and acceptor. We adopted ratios: 1:1, 2:3, 1:2. It has been achieved by the choice of colours of polygons with corresponding probabilities.

- The border between donor and acceptor may be active, i.e. such that the charge created after exciton's decay can get a proper electrode flowing by sequence of connected grains of donor/acceptor. Border can be also disactive when charges don't have such possibility (i.e. they are trapped in an island without contact with an electrode). Fig. 4 illustrates examples of active border (marked by green colour) and example of disactive border (blue colour).

- The length of the active border is measured and compared with length of border in 'comb architecture', treated as optimal.

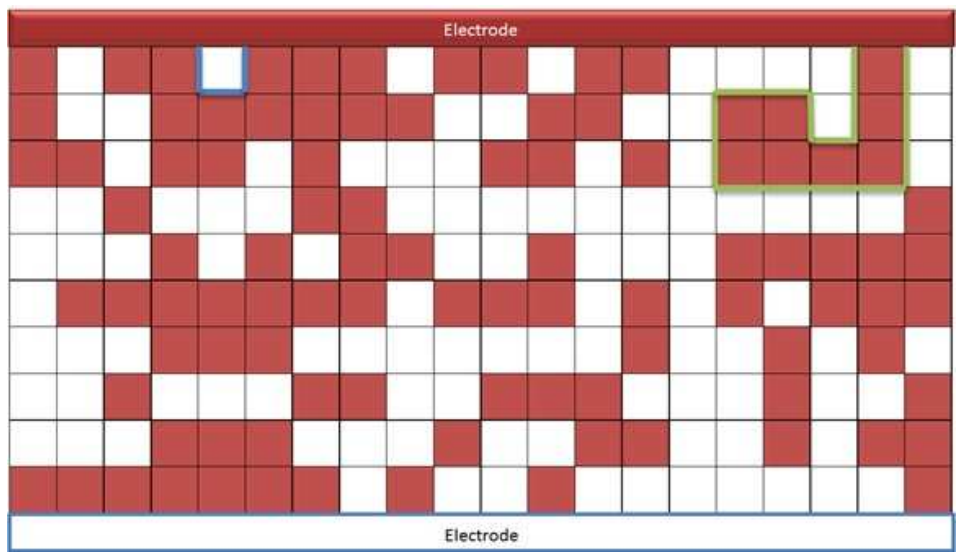

Fig. 4. The model of BHJ layer. The BHJ blend is formed as a random mixture of square grains of the donor and acceptor. The blue line is one of non-active borders, whereas the green line is one of active borders

## 2.3. Computational algorithm

Now, we present in more details an algorithm devoted to tasks above. We first present an algorithm for square lattice (Subsec. 2.3.1); in the next Subsec. 2.3.2 we describe an algorithm for honeycomb lattice.

### 2.3.1. Algorithm for the square lattice

We start from declaring a two-dimensional array indexed with pairs of positive integers (i, j). We anticipate that with two opposits sides of this array are connected electrodes and array is a model of an BHJ layer. In our simulations we used arrays of size ranging 10 x 10 till 20 x 200. Individual cells of declared array are coloured in one of two colours: white means that a given cell is filled by a grain of donor and red – by a grain of acceptor.

Parameters of programme are: vertical size V (thickness of the layer); horizontal size H; probability P of filling of given cell by grain of donor. The P value corresponds to ratio of donor and acceptor.

The main steps of the procedure are as follows.

**Step 1.** For each cell, a number from interval [0, 1] is cast. If this number is less than P, then a cell is filled with red colour. Otherwise it is filled with white colour. After the step 1, occupations of cells of the lattice by donor and acceptor have been determined.

**Step 2.** When coloring of cells of the array is finished, the programme again checks all the table to settle the quest of connection between „seeds" of donor and „seeds" of acceptor in cases when cells filled by this same colour contact only by corners. (Fig.5). For each cell programme checks it's colour and colours of neighbour cells. In case when cells filled by this same colour contact only by corners, a random numer from interval [0, 1] is generated. We assumed that in this case cells filled with this same colour are connected with probability 0,5. Obviously, when a pair of identical cells, for example white cells, is connected, then a pair of red cells is not connected. Information about connection or less of connection is scored up.

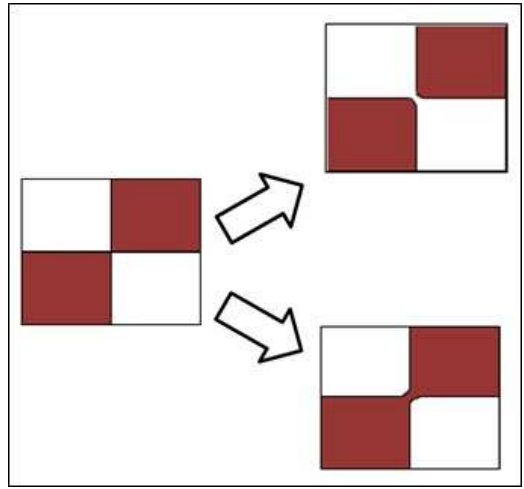

Fig. 5. In the case where squares of one type touch by vertex, we randomly choose connection between only 'donor' or only 'acceptor' squares

**Step 3.** When the quest of corner connections between cells is resolved, the programme enters into searching cells filled with red colour, which are connected with the „red" electrode (i.e. with an edge of the BHJ layer which – as assumed – collects the electrons) by a coherent path composed with the red cells. Analogically are searched the coherent paths composed by the white cells reaching a „white" electrode (i.e. an edge of the BHJ layer which collects holes).

**Step 4.** In this stage, there are determined these areas from which the charges can flow into an adequate electrode. In step 4 the length of a border between this areas is calculated. After this last step, the total length of the active border is known and an algorhitm is finishing one simulation.

Every four-step simulation is repeated N times (we took N=100). After that, standard statistical analysis of active border length is performed: We calculate the maximum, minimum, average, variance and standard deviation. Next, an quotient of average and the length of border between donor and acceptor in 'comb architecture' is appointed (in percent).

## 2.3.2. Algorithm for honeycomb lattice

The algorithm for the honeycomb lattice (Fig. 6) is very similar to this for the square lattice, so we only stress the differences.

-) There are some technical differences with declaration of tables, corresponding to two-dimensional array of hexagons forming a subset of hexagonal lattice.

-) The whole procedure for determination of active border between donor and acceptor is even simpler than its square-lattice analogon: Only three steps are involved, because donor and acceptor grains can touch only by edges and not by vertices, so the step 2 is skipped.

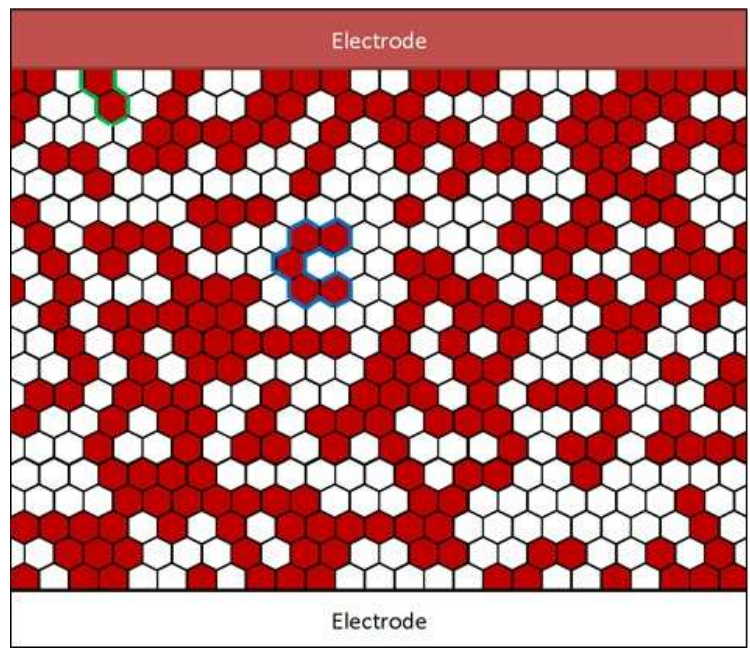

Fig. 6. The second model of BHJ layer. The BHJ blend is formed as a random mixture of hexagonal grains of the donor and acceptor. The blue line is one of non-active borders, whereas the green line is one of active borders

# 3. Results

## 3.1. Results for square lattice

We considered lattices of the following sizes: The lengths of shorter edge V were 10, 15 and 20 cells, which corresponded to thickness of the active layer being 100, 150 and 200 nm. The longer edge H ranged from 10 to 200 cells. In the Table 1 we present details of results obtained: minimal, maximal and average lengths of the active border between donor and acceptor, together with the optimal length (for the comb with teeth width equal to 1 cell) and the quotient Q of 'active' border length and ideal 'comb' length. The data in the table have been obtained for the proportion 1:1 of the donor and acceptor, and for N=100 independent casts of configurations.

We present results obtained on Fig.7. It is seen that the largest value of Q has been obtained for elongated samples. The value of Q grows with increasing elongation H/V and seems to stabilize, approaching certain limit value. The value of Q undergoes very little changes (less than 3%) when the elongation H/V is greater than 5. It is also clearly seen that the largest quotient Q has been achieved for smallest widths of active layer. The largest value of Q was 55 % and has been obtained for the sample 10 x 100.

We performed simulations of blends for other proportions of donor and acceptor. We took three values: 1:1, 2:3 and 1:2. We reproduce results only for values of Q; they are presented on Fig. 7.

| V x H | 10x10 | 20x20 | 10x30 | 10x50 | 15x45 | 20x60 | 15x100 | 20x100 |
|---|---|---|---|---|---|---|---|---|
| Minimal boundary length | 0 | 0 | 34 | 97 | 86 | 184 | 398 | 430 |
| Maximal boundary length | 58 | 217 | 204 | 358 | 405 | 766 | 945 | 1142 |
| Average boundary length $L\_A$ | 23,72 | 81,22 | 125,88 | 225,02 | 249,02 | 449,01 | 638,3 | 767,17 |
| Variance | 159,0723 | 2037,365 | 1371,379 | 2690,929 | 5445,05 | 15230,78 | 14245,57 | 27713,9 |
| Optimal length $L\_O$ | 82 | 362 | 262 | 442 | 617 | 1122 | 1387 | 1882 |
| Q = $L\_A/L\_O$ [%] | 29 | 22 | 48 | 51 | 40 | 40 | 46 | 41 |
| Standard deviation | 12,61239 | 45,13718 | 37,03214 | 51,87416 | 73,79058 | 123,413 | 119,3548 | 166,4749 |

Table 1. An example of detailed results of simulations for rectangular subsets of square lattice; V is vertical size (width), and H-horizontal size. The averages are taken from N=100 casts. Proportion of the donor and acceptor is 1:1.

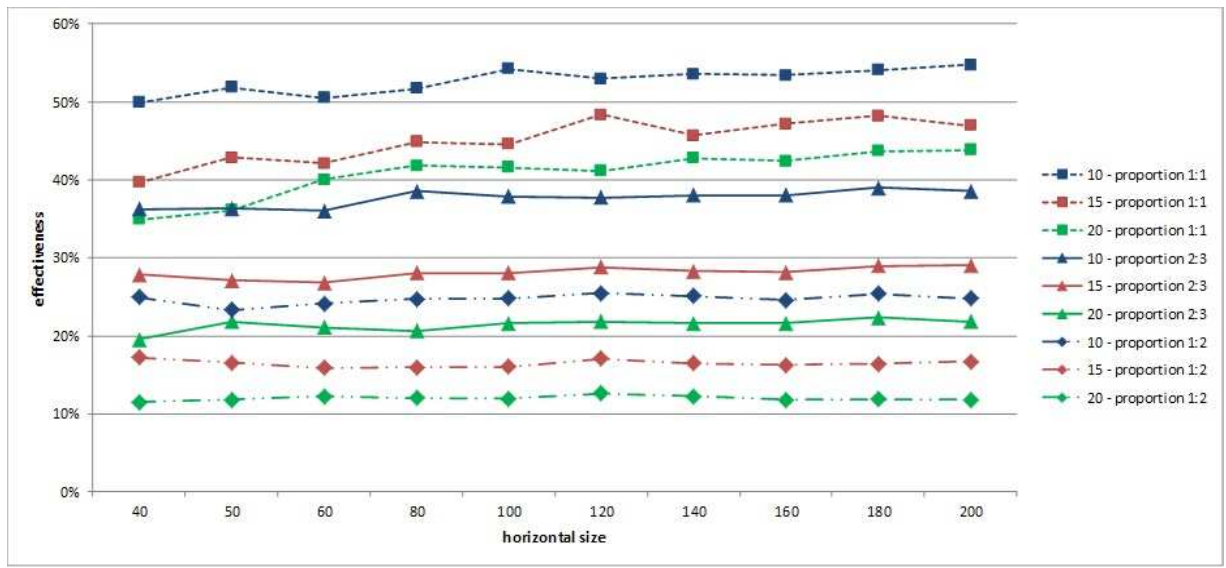

Fig. 7. Q values for the square lattice. Results for various horizontal and vertical sizes as well as various proportions of donor and acceptor are presented.

We observe that all plots of Q as a function of the horizontal size H possess very similar shape: Q value increases with increasing H and tend to certain limiting value, dependent of the vertical size V. The largest values of Q were observed for smallest values of V.

We also observe that the quotient Q decreases with deviation of the proportion D/A from 1. For the proportion D/A being 1:1, the largest value of Q was 55%; for D/A proportion equal to 2:3, the largest value of Q was 39%; and for D/A proportion 1:2, the largest value of Q was 27%.

### 3. 2. Results for honeycomb lattice

We considered lattices of sizes analogous as previously, i.e. for vertical size (width) being 10, 15 and 20 hexagons and horizontal size ranged from 10 to 200 hexagons. We took three proportions of donor and acceptor: 1:1, 2:3 and 1:2. In every case, the averages were calculated from N=100 independent casts, and the quotient Q was computed. We present results on Fig. 8.

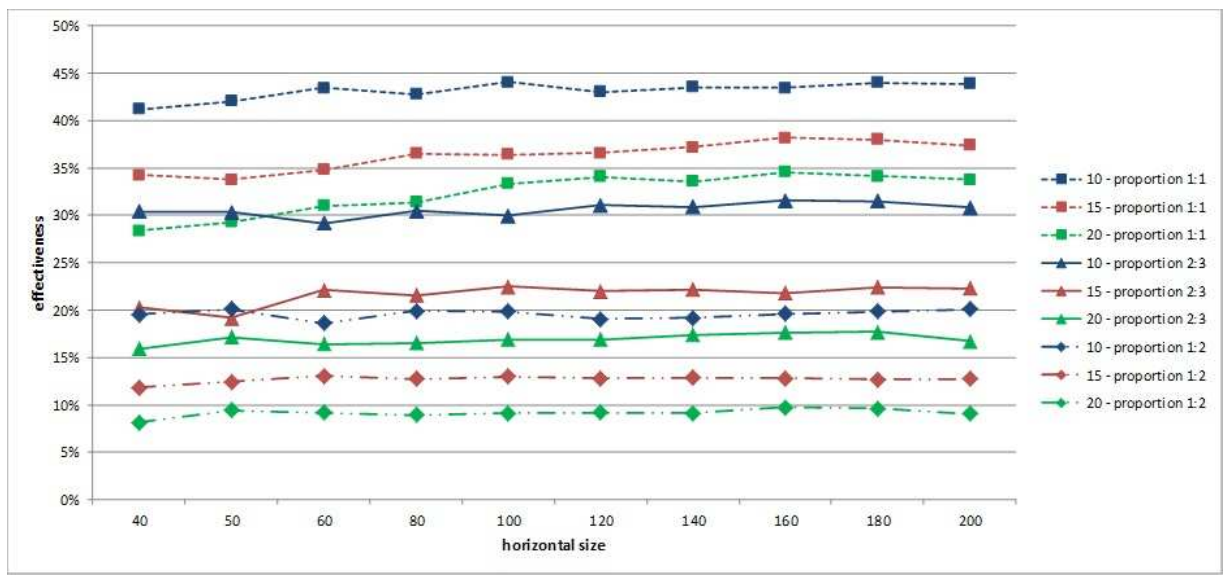

Fig. 8. Q values for the hexagonal lattice. Results for various horizontal and vertical sizes as well as various proportions of donor and acceptor are presented.

The general look of plots is very similar as in the square lattice case: The Q(H) functions are monotonically increasing and tend to certain limit value (depending of D/A proportion and vertical size). We observe that the values of Q(H) are *smaller* than for square lattice system. The difference is not large, but common for all D/A proportions and sizes. For instance, for proportion 1:1 and width 10, the maximal value of Q was 44% (55% for squares); for proportion 2:3 and width 10, the maximal value of Q was 31% (37% for squares); and for proportion 1:2 and width 10, maximal value of Q was 20% (27% for squares). For other widths we observe very similar interrelations.

**3.3. Summary of simulations**

The ultimate goal of our simulations was to find an answer to the question: Which is the length of the border between Donor and Acceptor in the BHJ architecture L_BHJ, compared with the length of the border in the ideal 'comb' architecture L_comb. The answer we have obtained is that *the length in the BHJ architecture is 2 or more smaller than in the 'comb' architecture.* In a more quantitative manner, the quotient of border lengths Q = L_BHJ/ L_comb was 0.2 – 0.55. The value of Q depend of the thickness of the active layer, proportion of the Donor and Acceptor and the shape of their grains. An immediate consequence of this opportunity is that the efficiency of the photovoltaic device could be about 2 times greater in the 'comb' architecture than in the BHJ architecture.

We have detected that in our model the most efficient proportion of the Donor and Acceptor was 1:1.

In the most casted configurations, we observed linking of two electrodes by the one kind of component (i. e. by a donor, or by an acceptor). Such a linking is present on exemplary casted configuration on Fig. 4. Presence of such linkings means that created electrons or holes can flow to both electrodes. Of course, it is unwanted process. The remedy to avoid it is well known – it is to use electron/hole blockades. Our results can be viewed as independent indication of necessity of using electron/hole blockades. Unfortunately, the presence of blockades makes the construction of solar cells more complicated

In principle, one could consider the statistics of such configurations (where electrodes are linked by one of species forming the active layer) by noticing that this phenomenon is certain kind of *percolation*. There are numerous results concerning this subject [17]. At present stage of research however we didn't analyse it in more details.

We observed dependence of the quotient Q of the shape of grains: For squares, the value of Q was systematically greater than the value for hexagons. It would be interesting to simulate systems with another shapes of grains (by looking at SEM results) and/or to allow presence of grains of different sizes. This will be one of our next attempts.

**4. Summary and conclusions**

The results obtained by us we treat as introductory ones. The basic opportunity is that we considered the two-dimensional model, whereas real systems are three-dimensional. However, we claim that consideration of two-dimensional model is instructive before the analysis of more involved three-dimensional case (like the relation between 2d and 3d Ising models for magnets). Our next tool is an analysis of three-dimensional model; we are actively pursuing in this direction.

The results presented above suggest that the BHJ architecture is non-optimal, and that the efficiency of the photovoltaic device in the optimal 'comb' architecture could be at least two times higher. But of course, the comb architecture is much more difficult to realize from technological point of view. The nanotubes appear here as natural candidates for combs.

The second direction of research aimed to enlarge the efficiency of photovoltaic devices could be to return to *layer* architecture. Devices constructed in a layer architecture exhibit lower (however not drastically) efficiency compared with those in BHJ architecture [5-8],[14]. To improve efficiency of 'layer' devices, one has to solve the main problem: *To find the substance(s), where the exciton diffusion length is comparable to the optical penetration length.* In more quantitative manner, typical value of optical penetration length is of the order 100 nm [14], so one should find the substance where the exciton diffusion length is of the order of 100 nm. It is very difficult task, as in the most of donors or acceptors used in photovoltaic devices the exciton diffusion length is of the order 10 nm [14]. But it seems that it is not hopeless, as there are known certain compounds (for instance, the anthracene) where the exciton diffusion length is about 100 nm! [14]. Unfortunately, the anthracene does not absorb the light in the visible range. To find the compound(s) absorbing the light in visible range and possessing the large exciton length is great challenge for material research.

**References**

[1] https://commons.wikimedia.org/wiki/FILE:Best_Research_Cell_Efficiencies.png

[2] A. J. Heeger : Semiconducting and Metallic Polymers: The Fourth Generation of Polymeric Materials (Nobel Lecture). *Angew. Chem. Int. Ed.* **2001**, 40, 2591-2611

[3] R. A. J. Janssen and J. Nelson: Factors Limiting Device Efficiency in Organic Photovoltaics. *Adv. Mater.* **2013**, *25*, 1847–1858

[4] B. P. Rand, H. Richter (eds.): Organic Solar Cells: Fundamentals, Devices, and Upscaling. CRC Press, Taylor & Francis Group 2014

[5] C. W. Tang: Two-Layer Organic Photovoltaic Cell. *Applied Physics Letters* 48, 183 (1986)

[6] Chu CW, Shao Y, Shrotriya V, Yang Y (2005) Efficient photovoltaic energy conversion in tetracene-C60 based heterojunctions. *Appl Phys. Lett,* **86,** 243506.

[7] Yuhki Terao, Hiroyuki Sasabe, and Chihaya Adachi: Correlation of hole mobility, exciton diffusion length, and solar cell characteristics in phthalocyanine/fullerene organic solar cells. *Applied Physics Letters* **90**, 193515 (2007)

[8] Xue J, Uchida S, Rand B. P., Forrest S.R.: Asymmetric tandem organic photovoltaic cells with hybrid planar-mixed molecular heterojunctions, *Appl. Phys. Lett.* **85**, 5757–5759 (2004) .

[9] G. Yu, A. J. Heeger, J. Appl. Phys. **78**, 4510 (1995).

[10] J. J. M. Halls, C. M. Walsh, N. C. Greenham, E. A. Marseglia, R. H.Friend, S. C. Moratti, A. B. Holmes, Nature **376**, 498 (1995).

[11] G. Yu, J. Gao, J. C. Hummelen, F. Wudl, A. J. Heeger: Polymer photovoltaic cells—enhanced efficiencies via a network of internal donor–acceptor heterojunctions. Science **270**, 1789 (1995).

[12] C. Y. Yang, A. J. Heeger: Synth. Met. **83**, 85 (1996).

[13] S. Günes, H. Neugebauer, N. S. Sariciftci, Conjugated polymer-based organic solar cells, Chem. Rev. **107**,1324-1338 (2007).

[14] R. R. Lunt and R. J. Holmes: *Small molecule and Vapor-Deposited Organic Photovoltaics.* In: B. P. Rand, H. Richter (eds.): Organic Solar Cells: Fundamentals, Devices, and Upscaling. CRC Press, Taylor & Francis Group 2014, Chap. 2.

[15] Qian Zhang, Bin Kan, Feng Liu, Guankui Long, Xiangjian Wan, Xiaoqing Chen, Yi Zuo, Wang Ni, Huijing Zhang, Miaomiao Li, Zhicheng Hu, Fei Huang, Yong Cao, Ziqi Liang,

Mingtao Zhang, Thomas P. Russell and Yongsheng Chen: Small-molecule solar cells with efficiency over 9%. Nature Photonics **9**, 35 – 41 (2015).

[16] Wei Chen, Feng Liu, Ondrej E. Dyck, Gerd Duscher, Huipeng Chen, Mark D. Dadmun, Wei You, Qiquan Qiao, Zhengguo Xiao, Jinsong Huang, Wei Ma, Jong K. Keum, Adam J. Rondinone, Karren L. More, and Jihua Chen: *Nanophase Separation in Organic Solar Cells.* In: Qiquan Qiao (ed.): Organic Solar Cells: Materials, Devices, Interfaces and Modeling. CRC Press, Taylor & Francis Group 2015, Chap. 9.

[17] D. Stauffer and A. Aharony: Introduction to Percolation Theory. CRC Press, Taylor & Francis 1994

[18] M. C. Scharber, N. S. Sariciftci: Efficiency of bulk-heterojunction solar cells. Progress in Polymer Science 38 (2013), 1929-1940.